\documentclass[aps,prl,twocolumn,showpacs,superscriptaddress,groupedaddress]{revtex4}  
\usepackage[colorlinks=true]{hyperref}
\usepackage{graphicx}  
\usepackage{dcolumn}   
\usepackage{bm}        
\usepackage{amssymb}   
\usepackage{xspace}
\usepackage{amsmath}
\usepackage{cancel}
\usepackage{multirow}
\usepackage[utf8]{inputenc}
\usepackage{pennames}
\usepackage{ptdr-definitions}
\usepackage{subfig}%
\let\subfigure\subfloat

\newcommand{\MR}{\ensuremath{M_\mathrm{R}}\xspace}

\newcommand{\Rtwo}{\ensuremath{\mathrm{R}^2}\xspace}

\newcommand{\met}{\ETm}
\newcommand{\mgaga}{\ensuremath{m_{\Pgg\Pgg}}\xspace}

\newcommand{\chizone}{\chiz_1}
\newcommand{\chiztwo}{\chiz_2}
\newcommand{\sbottom}{\PSQb}

\begin{document}

\widetext


\title{Squark-mediated Higgs+jets production at the LHC}
\author{J.~Duarte\footnote{jduarte@caltech.edu; now at Fermi National Accelerator Laboratory.}} \affiliation{Caltech, 1200 E California  Blvd, Pasadena, CA 91125 USA}
\author{C.~Pe\~{n}a} \affiliation{Caltech, 1200 E California  Blvd, Pasadena, CA 91125 USA}
\author{A. Wang\footnote{Now at Harvard University.}} \affiliation{Caltech, 1200 E California  Blvd, Pasadena, CA 91125 USA}
\author{M.~Pierini} \affiliation{CERN, Rte de Meyrin, Geneve, 1211 Switzerland}
\author{M.~Spiropulu} \affiliation{Caltech, 1200 E California  Blvd, Pasadena, CA 91125 USA}

\date{\today}

\begin{abstract}
We investigate possible scenarios of light-squark production at the
LHC as a new mechanism to produce Higgs bosons in association with
jets. The study is  motivated by the SUSY search for $\PH+$jets events,
performed by the CMS collaboration on $\sqrt{s}=8, 13 \TeV$ data  using  the
\textit{razor} variables. Two simplified models are proposed to interpret  the observations in this search. The constraint from Run I and
the implications for Run II and beyond are discussed. 
\end{abstract}

\maketitle
%
\section{Introduction}
 
The ATLAS and CMS collaborations has searched intensively  for SUSY production in the data collected at a center-of-mass energy $\sqrt{s}=8 \TeV$ in 2012. A large part of the searches focused on SUSY models with conserved R-parity, for which the lightest SUSY particle (LSP) is stable. The LHC is particularly sensitive to the production of SUSY partners charged under QCD (squarks and gluinos), given the  dominant hadroproduction cross section in proton-proton collisions. Following  the stringent bounds on generic SUSY models obtained  with $\sqrt{s}=7 \TeV$ data, ATLAS and CMS moved the focus of their SUSY searches to the so-called \textit{natural} SUSY models~\cite{Papucci:2011wy}. In its minimal realization, a natural SUSY spectrum is composed of the minimum set of SUSY partners needed to protect the mass of the Higgs ($\PH$) boson from quantum corrections: a gluino, one bottom squark, two top squarks, and three higgsinos (two neutral and one charged). This SUSY scenario results in events with multiple top and bottom quarks, produced in association with missing transverse energy ~\met. No evidence for the production of such particles has been  found, pushing the allowed mass range for gluinos and top
squarks above $\sim 1600 \GeV$ and $\sim 700 \GeV$, respectively, for a low-mass neutralino LSP and largely independent of the top squark and gluino branching ratios (see for instance Ref.~\cite{razor8TeV,CMS-PAS-SUS-15-004}). 

In a few cases, a data yield above the expected background was observed for certain signal regions, for example, in the case of the \emph{edge} dilepton analysis by CMS~\cite{CMSedge} and the SUSY search in $\PZ+$jets events by ATLAS~\cite{ATLASZpeak}. These excesses correspond to, respectively, $\sim 2.4\sigma$ and $\sim 3.0\sigma$ of
local significance, which are reduced after accounting for the look-elsewhere effect (LEE). Several interpretations of these results
were given in the literature~\cite{Theory1,Theory2,Theory3,Theory4,Theory5,Theory6}, mainly related to the electroweak production of SUSY particles with long decay chains. 

Here we discuss the re-interpretation of the  search
for electroweak SUSY partners in $\PH (\Pgg\Pgg)+ \geq 1$~jet events by the CMS
collaboration performed at $8\TeV$~\cite{RazorHgaga}. The analysis uses the diphoton
invariant mass \mgaga to select events with a $\PH$-like
candidate. The non-resonant (mostly QCD
diphoton production) and resonant (standard model $\PH(\Pgg\Pgg)$
production) backgrounds are estimated using the \mgaga sidebands in data
and the Monte Carlo simulation, respectively. The background prediction is performed as a
function of the razor variables $\MR$ and $\Rtwo$ in five mutually
exclusive \emph{boxes}, targeting different final states:
high-$\pt$ $\PH (\Pgg\Pgg)$ (\texttt{HighPt} box), $\PH
(\Pgg\Pgg)+\PH (\bbbar)$ (\texttt{Hbb} box), $\PH
(\Pgg\Pgg)+\PZ (\bbbar)$ (\texttt{Zbb} box), and low-$\pt$ $\PH
(\Pgg\Pgg)$ with high- and low-resolution photons
(\texttt{HighRes} and \texttt{LowRes} boxes, respectively). Five events are
observed in one ($\MR$, $\Rtwo$) bin of the \texttt{HighRes} box, compared
to less than one expected background event. This corresponds to a
local significance of $2.9\sigma$, reduced to $1.6\sigma$ after the
LEE. 

In this paper, we propose and study a new interpretation of this search in
terms of SUSY models with light quarks. We emulate this CMS analysis
to derive bounds on squark production. Since the analysis does not
require or veto jets originating from \cPqb-quarks (\cPqb-jets), the results
apply to bottom-squark production in natural SUSY models. 

Recently, an updated search was performed with data collected at
$13\TeV$~\cite{CMS-PAS-SUS-16-012}. One of the models proposed during the studies presented in this paper (model B)  was also used for the interpretation of the results.

\section{Benchmark signal models}
\label{sec:models}

We consider two simplified models with bottom squark pair production, both
resulting in a $\PH$+jets final state.

In the first model, hereafter referred to as model A, we consider the
asymmetric production of a $\sbottom_2
\sbottom_1$ pair, where $\sbottom_2$ and $\sbottom_1$ are the heaviest
and the lightest bottom squarks, respectively. The $\sbottom_2$ decays
to $\cPqb \chiztwo$, with $\chiztwo \to \PH \chizone$. The lightest neutralino $\chizone$ is
assumed to be the LSP. The $\sbottom_1$, close in mass to the LSP,
decays to $\cPqb \chizone$. All the other SUSY partners are assumed to be
too heavy to be produced at the LHC and are ignored in this
analysis. This model represents a new mechanism for the production of
$\PH+2\cPqb\textrm{-}\mathrm{jets}+\mathrm{invisible}$, with one of
the associated \cPqb-jets typically having low momentum.

In the second model, hereafter referred to as model B~\cite{annthesis}, two bottom
squarks $\sbottom_1\sbottom_1$ are produced, each decaying as
$\sbottom_1 \to \cPqb \chiztwo$. The $\chiztwo$ then decays to $\PH
\chizone$, the $\chizone$ being the LSP. As for model A, the other SUSY
partners are ignored. This simplified model corresponds to a final
state consisting of $2\PH+2\cPqb\textrm{-}\mathrm{jets}+\mathrm{invisible}$.

\begin{figure}[htb]
\includegraphics[width=0.23\textwidth,viewport=250 100 800 700,clip=true]{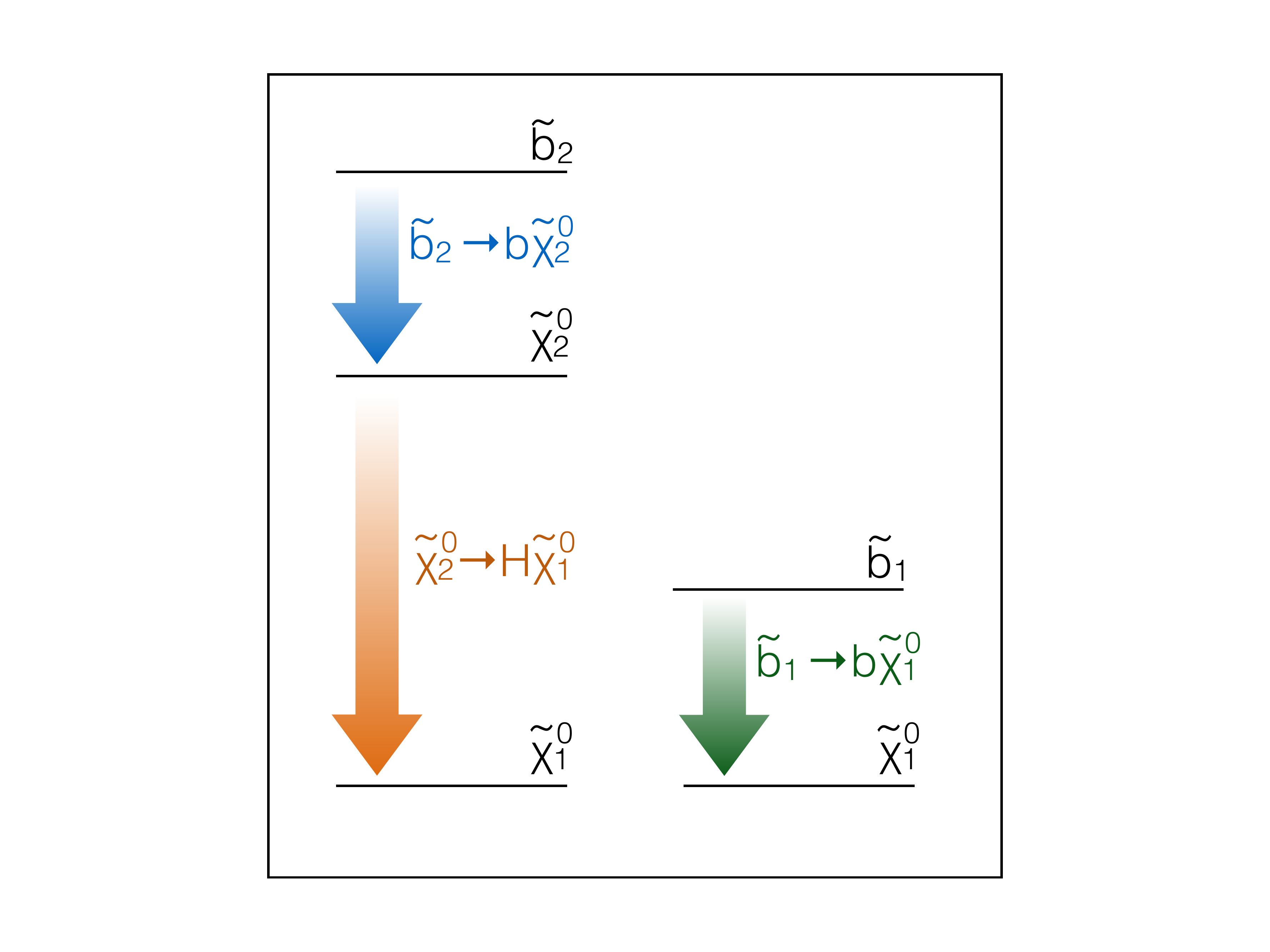}
\includegraphics[width=0.23\textwidth,viewport=250 100 800 700,clip=true]{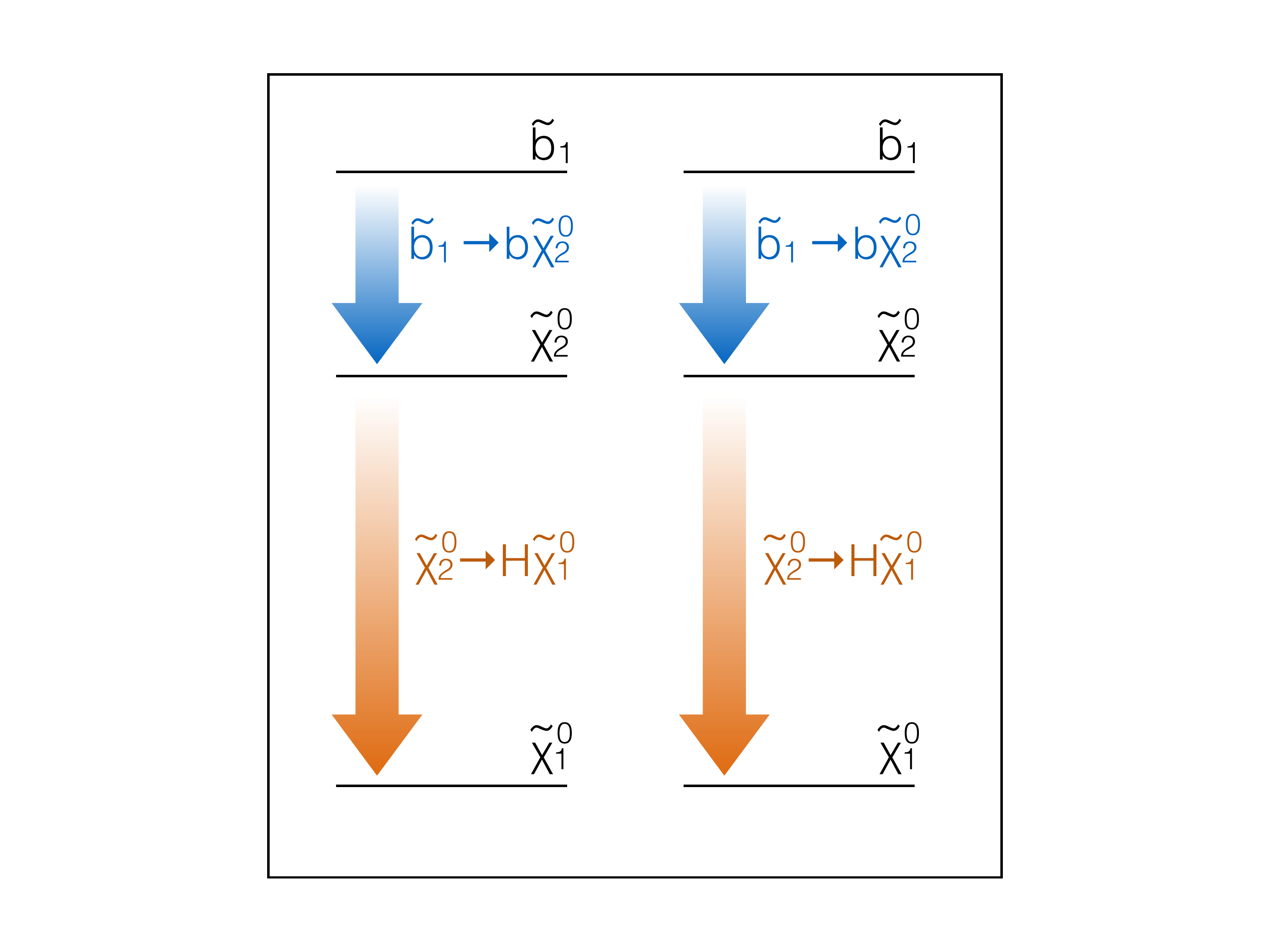}
\includegraphics[width=0.23\textwidth]{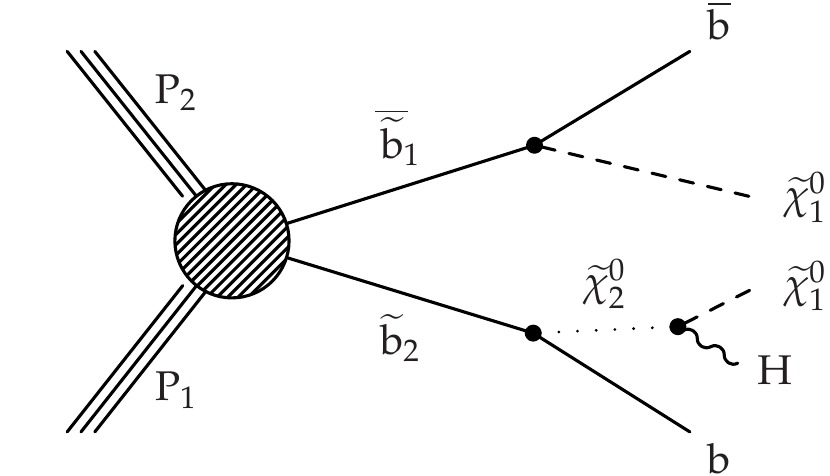}
\includegraphics[width=0.23\textwidth]{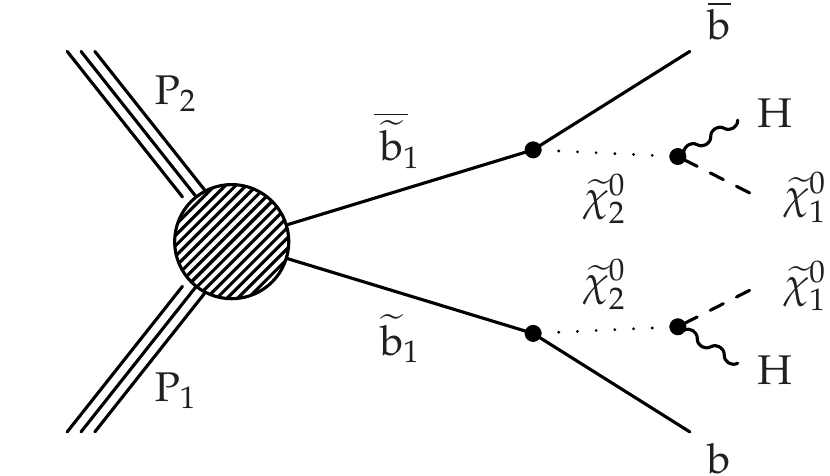}
\caption{\label{fig:simplifiedModels} Pictorial representation of the
  decay chains and event topologies associated with model A (left) and model B (right), as described in the text.}
\end{figure}

The mass spectrum for each model is shown in
Fig.~\ref{fig:simplifiedModels}. We fix the $\chiztwo$ and $\chizone$ masses to 230\GeV
and 100\GeV, respectively.  In model A,
we fix the $\sbottom_1$ mass to 130\GeV as varying its mass
in between the limits of the $\chizone$ and $\chiztwo$ masses has little effect.
Finally, we scan the $\sbottom_2$ ($\sbottom_1$) mass between 250\GeV
and 800\GeV for model A (B). These assumptions do not limit the
conclusions derived on the squark production cross section. In fact,
the analysis is sensitive to mass differences and not to the absolute
mass of SUSY partners. On the other hand, the chosen LSP and NLSP
masses does play a role when the cross section limits are
translated in terms of mass exclusion bounds.

\section{Event generation and detector simulation}
\label{sec:gensim}
The study is performed using samples of Monte Carlo events. The event generation is performed in \textsc{PYTHIA}~8.210~\cite{Pythia64,Pythia82}.
The default parton density function set is \textsc{NNPDF}~2.3 QCD+QED
LO (with $\alpha_s(m_\PZ) =
0.130$)~\cite{NNPDF1,NNPDF2,NNPDF3}. Fast simulation of the 
detector is performed in \textsc{Delphes}~3.3.2~\cite{Delphes3}. The
default description of CMS as provided in the release is used, except
for a modification to the photon isolation and efficiency,
described in the next section. Jet clustering is performed using
\FASTJET~3.1.3~\cite{fastjet}. As in CMS, the anti-$\kt$ jet
clustering algorithm is used with jet-size parameter $R=0.5$~\cite{antikt}.

\section{Emulation of the CMS search}\label{sec:emulation}

The emulated event selection is summarized as follows,
\begin{itemize}
\item Events with two isolated photons with $\pt>25$\GeV and
  $|\eta|<1.44$ are selected. As in Ref.~\cite{CMSPhoton}, the photon
  isolation variables, $I_{\gamma}$, $I_{\mathrm{n}}$, and $I_{\pi}$, are
  computed by summing the transverse momenta of photons, neutral
  hadrons, and charged hadrons, respectively, inside an isolation
  cone of radius $\Delta R=0.3$ around the selected photon. The photon
  isolation requirements on these variables
  are shown in Tab~\ref{tab:isolation}. An additional photon selection
  efficiency is applied in \textsc{Delphes} such that isolated photons with $\pt<10$\GeV ($\pt\geq10$\GeV) are
  randomly selected with 94\% (98\%) efficiency.
\begin{table}\centering
\caption{\label{tab:isolation}Photon isolation requirements, as in
  Ref~\cite{CMSPhoton}. The photon isolation variables, $I_{\gamma}$, $I_{\mathrm{n}}$, and $I_{\pi}$, are
  computed by summing the transverse momenta of photons, neutral
  hadrons, and charged hadrons, respectively, inside an isolation
  cone of radius $\Delta R=0.3$ around the selected photon.}
\begin{tabular}{lc|r}\hline\hline
 \multirow{2}{*}{$I_{\gamma}$} & barrel & $1.3\GeV + 0.005\pt^{\gamma}$\\
 & endcap & -- \\\hline
 \multirow{2}{*}{$I_{\mathrm{n}}$} & barrel & $3.5\GeV+ 0.04\pt^{\gamma}$\\
 & endcap &  $2.9\GeV+ 0.04\pt^{\gamma}$ \\\hline
 \multirow{2}{*}{$I_{\pi}$} & barrel & $2.6\GeV$\\
 & endcap &  $2.3\GeV$ \\\hline\hline
\end{tabular}
\end{table}
\item Events with one $\PH$ candidate with $\pt>20$ GeV are selected. A pair of selected
  photons is considered an $\PH$ candidate if at
  least one photon has $\pt>40$\GeV and the diphoton mass
  $m_{\Pgg\Pgg}>100$\GeV. If the event contains more than one $\PH$ candidate,
  the one with the highest scalar sum $\pt$ of the two photons is selected. 
\item Jets are reconstructed using the \FASTJET~\cite{fastjet} implementation
  of the anti-$\kt$~\cite{antikt} algorithm with jet radius parameter $R=0.5$.
\item Events with at least one jet with $\pt>30$ GeV and $|\eta|<3.0$
  are selected.
\item An emulation of the ``medium'' requirement (mistag probability of
  1\% and \cPqb-tag efficiency of $\sim 68\%$) of the combined secondary vertex (CSV) \cPqb-tagging
  algorithm is used to identify \cPqb-jets~\cite{btag8TeV}.
\item A $\bbbar$ candidate pair is identified if both jets satisfy the medium requirement of
  the \cPqb-tagging algorithm (note: the CMS analysis requires only one to
  satisfy the medium requirement, while both are required to satisfy
  the loose requirement).
\item The $\bbbar$ candidate pair with the mass closest to 125\GeV or 91.2\GeV is chosen as the $\PH\to
  \bbbar$ or $\PZ\to \bbbar$ candidate, respectively.
\item The razor variable \MR, calculated from two megajets~\cite{razorPRD} is required to be greater than
  $150\GeV$. All possible combinations of the reconstructed jets and
the $\PH(\Pgg\Pgg)$ candidate are clustered to form megajets. The pair of megajets that
minimizes the sum in quadrature of the invariant masses of the two megajets is selected.
\end{itemize}

After this baseline selection, events are categorized according to the
following requirements,
\begin{itemize}
\item \texttt{HighPt}: all events with an $\PH\to\Pgg\Pgg$ candidate
  with $p_{T}>110$\GeV. 
\item \texttt{Hbb}: remaining events with a $\PH\to \bbbar$ candidate
  with mass $110\geq m_{\bbbar}\geq 140$\GeV. 
\item \texttt{Zbb}: remaining events with a $\PZ\to \bbbar$ candidate
  with mass $76\geq m_{\bbbar}\geq 106$\GeV. 
\item \texttt{HighRes}: 70\% of remaining events after the
  \texttt{Zbb} selection (emulating the efficiency of the ``high-resolution photon'' selection).
\item \texttt{LowRes}: all remaining events. 
\end{itemize}
We assume the breakdown of events between the \texttt{HighRes} box and \texttt{LowRes}
box is 70\%-to-30\% after the \texttt{Zbb} selection. This
is based on the following observations:  (i) CMS categorizes events in the \texttt{HighRes} box if
both photons in the event satisfy $\sigma_E/E < 0.015$, where $\sigma_E/E$ is the estimated
relative energy resolution, and categorizes events in the
\texttt{LowRes} box otherwise, (ii) CMS observes a similar
70\%-to-30\% breakdown for both SM Higgs production and
electroweak SUSY processes in Monte Carlo
simulation~\cite{RazorHgaga}, and (iii) we expect this breakdown to be
model-independent assuming both photons are real and come from the
decay of a Higgs boson, as it is based on the properties of such photons
detected in CMS and not on the details of the model.

Finally, the search region selection is as follows,
\begin{itemize}
\item The search region in the $m_{\Pgg\Pgg}$ distribution is
    defined by $(125 - 2\sigma_{\mathrm{eff}},
    126+2\sigma_{\mathrm{eff}})$ in each event category, where
    $\sigma_{\mathrm{eff}}$ is defined such that $\sim68\%$ of Higgs
    boson events fall in an interval of $\pm\sigma_{\mathrm{eff}}$
    around the nominal $m_\PH$ value. Following this procedure using
    our generated and simulated signal samples, we derive
    $\sigma_{\mathrm{eff}}$ to be 3.8\GeV in the \texttt{HighPt} box
    and 2.2\GeV in the \texttt{HighRes} and \texttt{LowRes}
    boxes. For the \texttt{Hbb} and \texttt{Zbb} boxes, due to the low
    number of selected signal events, we use the overall average value
    of 2.8\GeV. 
   
\end{itemize}
We note that these $\sigma_{\mathrm{eff}}$ values are larger than the
    corresponding ones in Ref.~\cite{RazorHgaga}. This is due to the
    larger width observed for the diphoton mass distribution in Higgs
    boson events simulated and reconstructed with \textsc{Delphes},
    compared to official CMS software. This implies the effective
    diphoton mass resolution when using \textsc{Delphes} is larger than in the
    real CMS detector. We attempt to account for this with a
    modification explained in Sec.~\ref{sec:validation}.

\section{Bayesian Statistical Interpretation}\label{sec:bayes}

We model the likelihood according to a Poisson density,
considering the expected background yield (with associated
uncertainty), the expected signal yield (for a given signal cross
section), and the observed yield. The background uncertainty is modeled with a gamma density. The
background yields and the corresponding uncertainties are taken from the tables provided
in Ref.~\cite{RazorHgaga}. To take into account systematic
uncertainties on the signal, we assign a 30\% uncertainty (assuming a
log-normal density) on the signal strength, a multiplicative
factor modifying the signal cross section. We then derive the
posterior density for the signal cross section $\sigma$ as:
\begin{equation}
p(\sigma|\mathrm{data}) \propto \mathcal L(\mathrm{data} |\sigma)p_0(\sigma)~,
\label{eqn:posterior}
\end{equation}
where $\mathcal L(\mathrm{data} |\sigma)$ is the likelihood and $p_0(\sigma)$ is the prior density taken to be
uniform. The likelihood is then
\begin{align}
\mathcal L(\mathrm{data} |\sigma)
  &=\int_{0}^{\infty}\mathrm{d}\mu~\mathrm{Ln}(\mu|\bar\mu,\delta\mu)\\
&\times\prod_{i=0}^{n_{\mathrm{bins}}}\int_0^{\infty} \mathrm{d}b_i
   \mathrm{Poisson}(n_i|L\mu\sigma\epsilon_i+ b_i)\nonumber\\
&\times\Gamma(b_i|\bar{b}_i,\delta b_i)~,
\label{eqn:likelihood}
\end{align}
where the product runs over the number of bins $n_{\mathrm{bins}}$; $n_i$ is the
observed yield in the $i^{\mathrm{th}}$ bin, $L$ is the integrated
luminosity, $b_i$ is the assumed value of the background yield in the
$i^{\mathrm{th}}$ bin and $\bar{b}_i\pm \delta b_i$ is its expected value
and the associated uncertainty; $\epsilon_i$ is the nominal value
of the signal efficiency times acceptance in the $i^{\mathrm{th}}$ bin; $\mu$ is the
signal strength, a nuisance parameter modifying the signal cross section
(nominally equal to $\bar\mu=1$ with a $\delta\mu=30\%$ uncertainty);
$\mathrm{Ln}(x|m,\delta)$ is the log-normal
distribution for $x$, parameterized such that $\mathrm{log}(m)$ is the
mean and $\mathrm{log}(1+m\delta)$ is the standard deviation of the
log of the distribution; $\Gamma(x|m,\delta)$ is the gamma
distribution for $x$, parameterized such that $m$ is the
mode and $\delta^2$ is the variance of the distribution. The 95\%
credibility level (CL) upper limit on the
signal cross section $\sigma_{\mathrm{up}}$ is obtained from the
posterior, such that 
\begin{equation}
\frac{\int_0^{\sigma_{\mathrm{up}}}\mathrm{d}\sigma~ p(\sigma|\mathrm{data})}{\int_0^{\infty}\mathrm{d}\sigma~ p(\sigma|\mathrm{data})} = 0.95~.
\end{equation}

We also utilize a signal significance measure defined by
\begin{align}
Z(\sigma) &= \mathrm{sign}[\log B_{10}(\mathrm{data},\sigma)]\sqrt{2|\log B_{10}(\mathrm{data},\sigma)|}~,
\label{eqn:zSig}
\end{align}
where 
\begin{align}
B_{10}(\mathrm{data},\sigma) &= \frac{\mathcal L(\mathrm{data}
  |\sigma,H_1)}{\mathcal L(\mathrm{data}
  |H_0)}
\label{eqn:localBayes}
\end{align}
is the \emph{local} Bayes factor for the data for a given signal cross
section $\sigma$, and $\mathcal L(\mathrm{data}
  |\sigma,H_1)$ and $\mathcal L(\mathrm{data}
  |H_0)$ are the likelihoods for the signal-plus-background ($H_1$) and
  background-only ($H_0$) hypotheses, respectively. As described in
  Ref.~\cite{CMS-PAS-SUS-15-010}, this
  measured is a signed Bayesian analog of the frequentist ``n-sigma.''
  For each signal model with specified masses, we scan the signal
  cross section $\sigma$ to find the maximum significance,
  which occurs at the mode of the posterior.

\section{Correction and Validation}\label{sec:validation}

As discussed above, we find differences in the
performance of the emulated CMS detector and the real CMS detector,
e.g. the larger diphoton mass resolution. To take into account this
and other differences in the detector simulation and reconstruction performed by
\textsc{Delphes} and official CMS software, we conservatively double the
background uncertainties in each bin reported by CMS in Ref.~\cite{RazorHgaga} when evaluating the likelihood in
Eqn.~\ref{eqn:likelihood}. We find this conservative approach better
reproduces the observed and expected limits on a benchmark simplified
model.

To validate our emulation result, we produced 95\% CL
limits on the production cross section of an electroweak simplified
model of $\chipm_1\chiz_2$ production, followed by
the decays $\chipm_1\to \PW^{\pm}\chiz_1$,
$\chiz_2\to \PH\chiz_1$. For this model, CMS provided the 95\%
confidence level upper limits on the cross section assuming an LSP mass of
$m_{\chiz_1}=1\GeV$ and equal chargino and second neutralino
masses, $m_{\chipm_1}=m_{\chiz_2}$. 

\begin{figure}[htb]\centering
\includegraphics[width=0.45\textwidth]{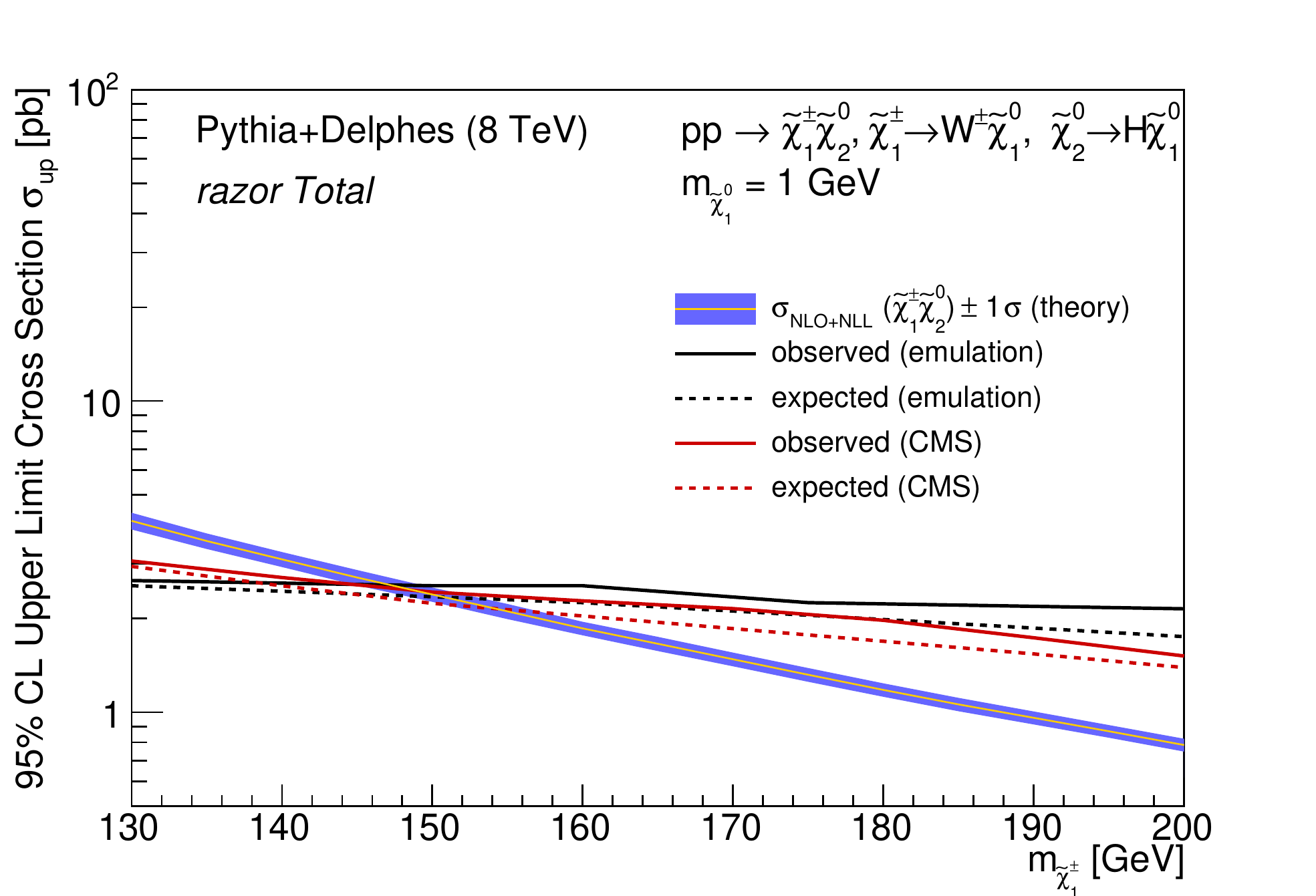}
\caption{\label{fig:TChiwh1dLimit} Comparison between the CMS
 result (red) and our emulation (black). Note, this scan assumes
 $m_{\chiz_1}=1$\GeV and $m_{\chipm_1}=m_{\chiz_2}$.}
\end{figure}

The comparison between our result and the CMS result for this model is shown in
figure~\ref{fig:TChiwh1dLimit} as a function of $m_{\chipm_1}$.

\section{Results}
\label{sec:results}

\begin{figure}[htb]\centering
\begin{tabular}{c}
\subfigure{\includegraphics[width=0.45\textwidth]{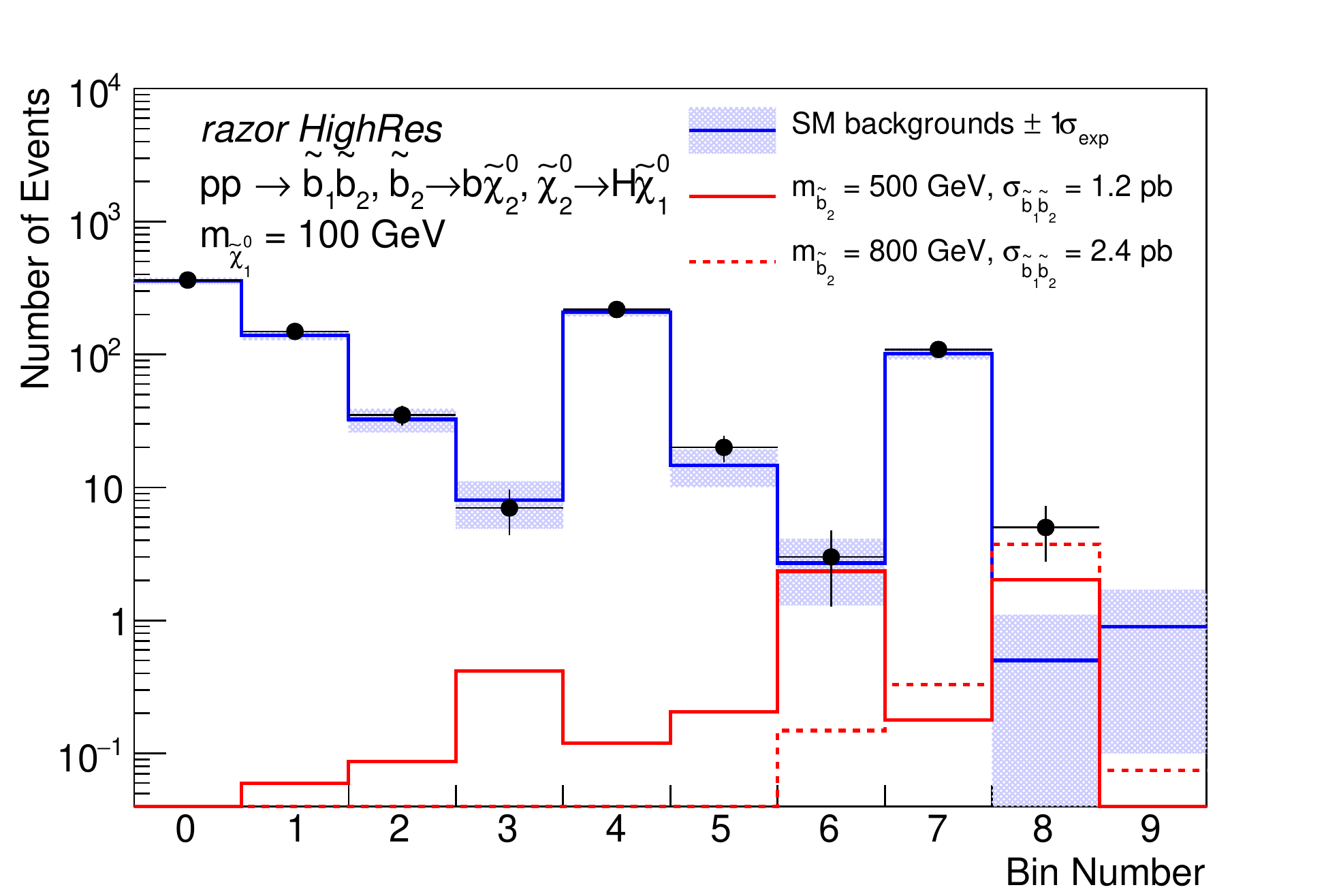}}\\
\subfigure{\includegraphics[width=0.45\textwidth]{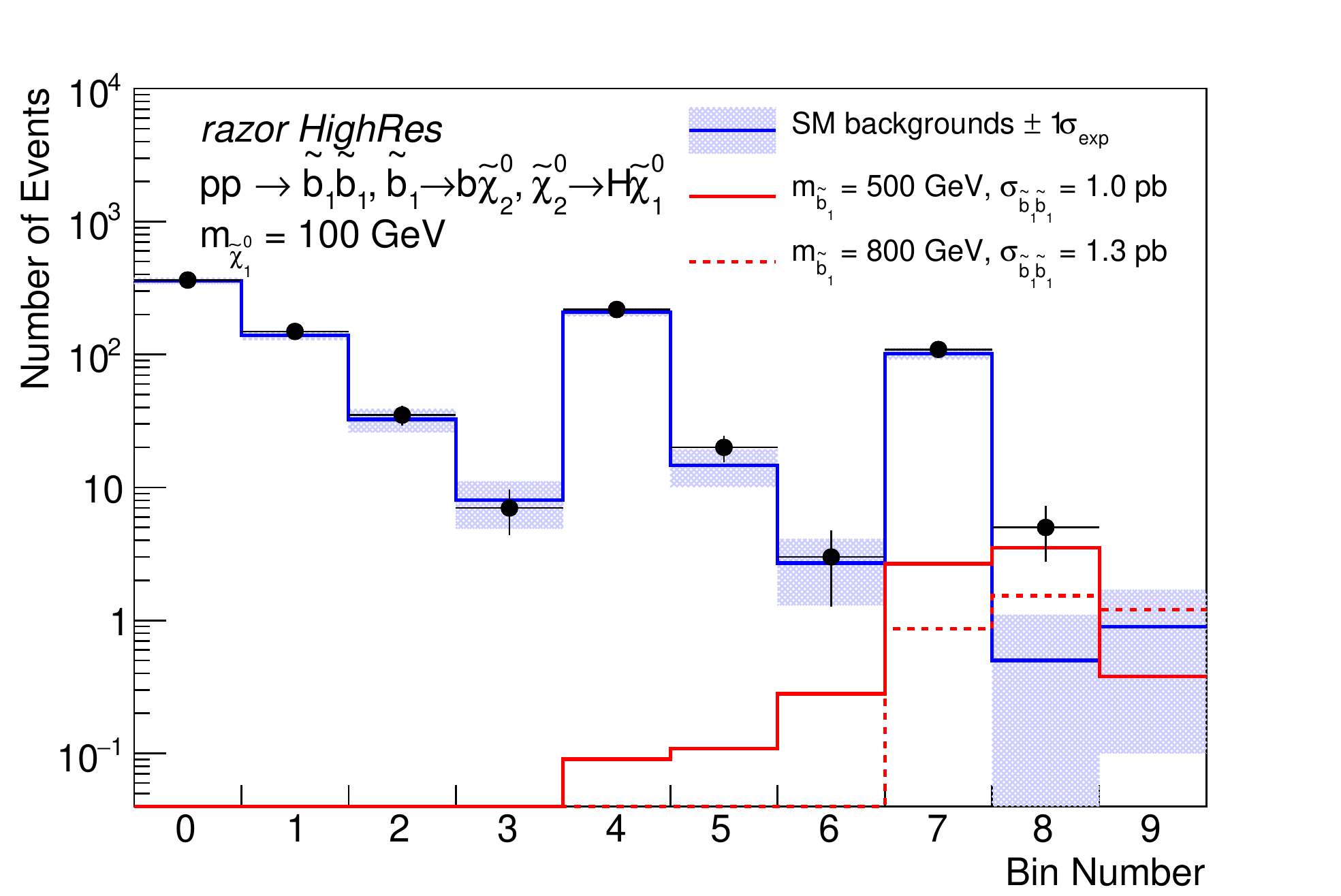}}
\end{tabular}
\caption{\label{fig:T21bHT2bHExpObs500800} (Top) The expected background and
  uncertainty (multiplied by a factor of two as explained in the text) compared to the best-fit signal distribution in the \texttt{HighRes} box for two particular
  mass points, $m_{\sbottom_2}=500\GeV$ and $m_{\sbottom_2}=800\GeV$,
  in model A. (Bottom) The expected background and
  uncertainty (multiplied by a factor of two as explained in the text) compared to the best-fit signal distribution in the \texttt{HighRes} box for two particular
  mass points, $m_{\sbottom_1}=500\GeV$ and $m_{\sbottom_1}=800\GeV$,
  in model B. The bin numbers correspond to the order of the signal regions in the yield tables in Ref.~\cite{RazorHgaga} and are reproduced in Tab.~\ref{tab:bins}.}
\end{figure}

\begin{figure}[htb]\centering
\subfigure{\includegraphics[width=0.45\textwidth]{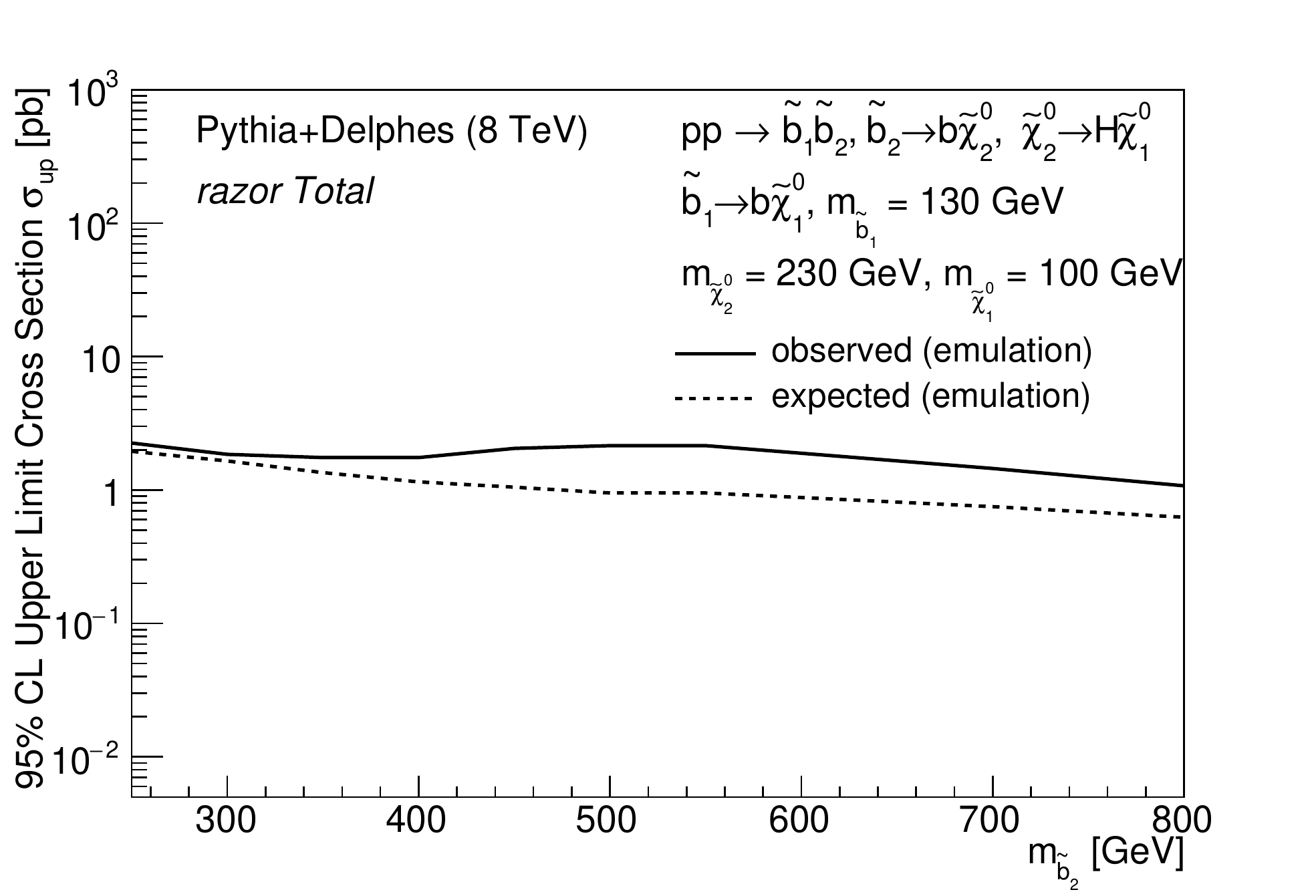}}\\
\subfigure{\includegraphics[width=0.45\textwidth]{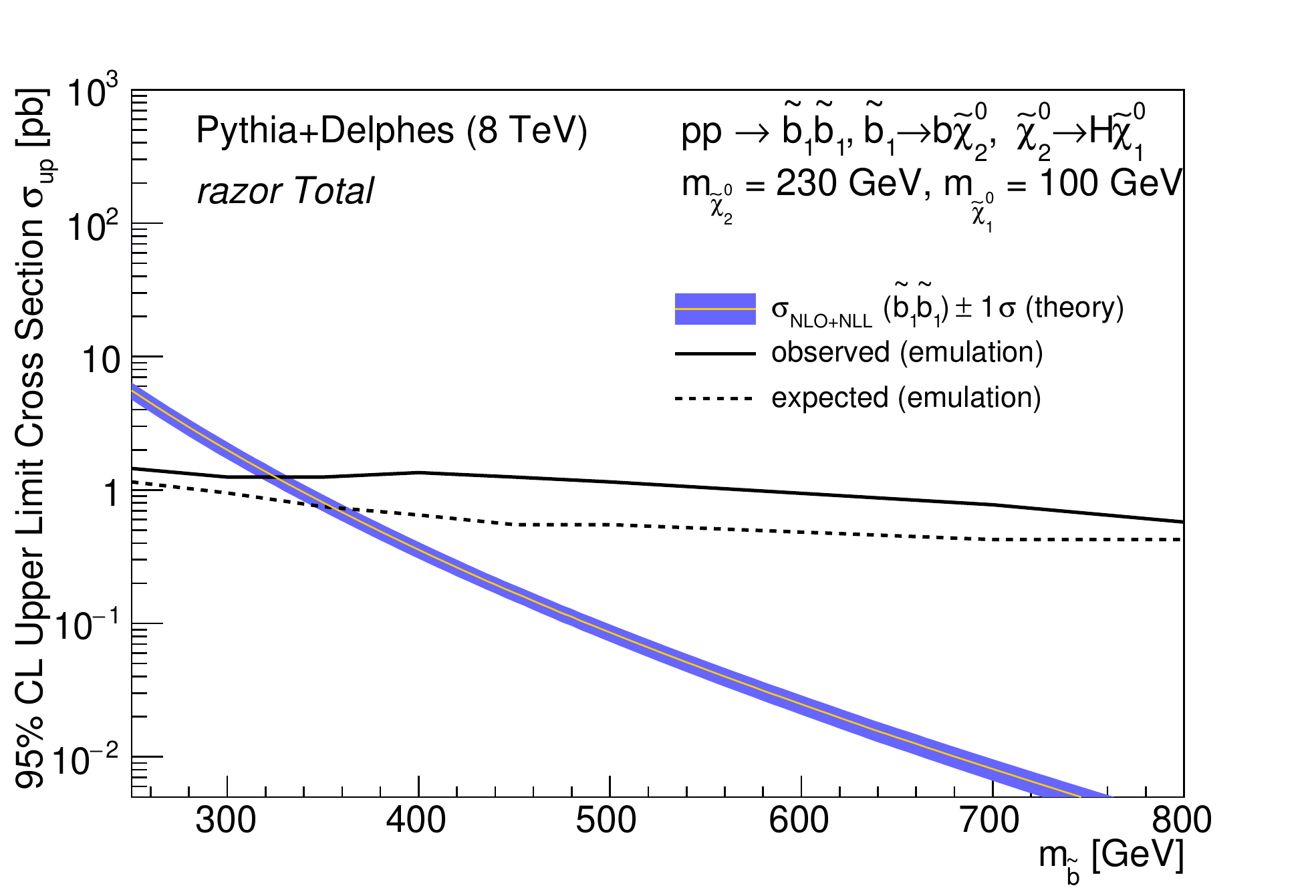}}
\caption{\label{fig:T21bHT2bH1dLimit} (Top) The 95\% CL upper limit on the
  cross section on $\sbottom_1\sbottom_2$ production in model A as a function of $m_{\sbottom_2}$ (black). (Bottom) The 95\% CL upper limit on the
  cross section on $\sbottom_1\sbottom_1$ production in model B as a function of $m_{\sbottom_1}$ (black) compared
  to the NLO+NLL predicted cross section (yellow). Note, these scans assume
  $m_{\chiz_1}=100\GeV$, $m_{\chiz_2}=230\GeV$, and for model A $m_{\sbottom_1}=130$\GeV. }
\end{figure}

\begin{figure}[htb]\centering
\subfigure{\includegraphics[width=0.45\textwidth]{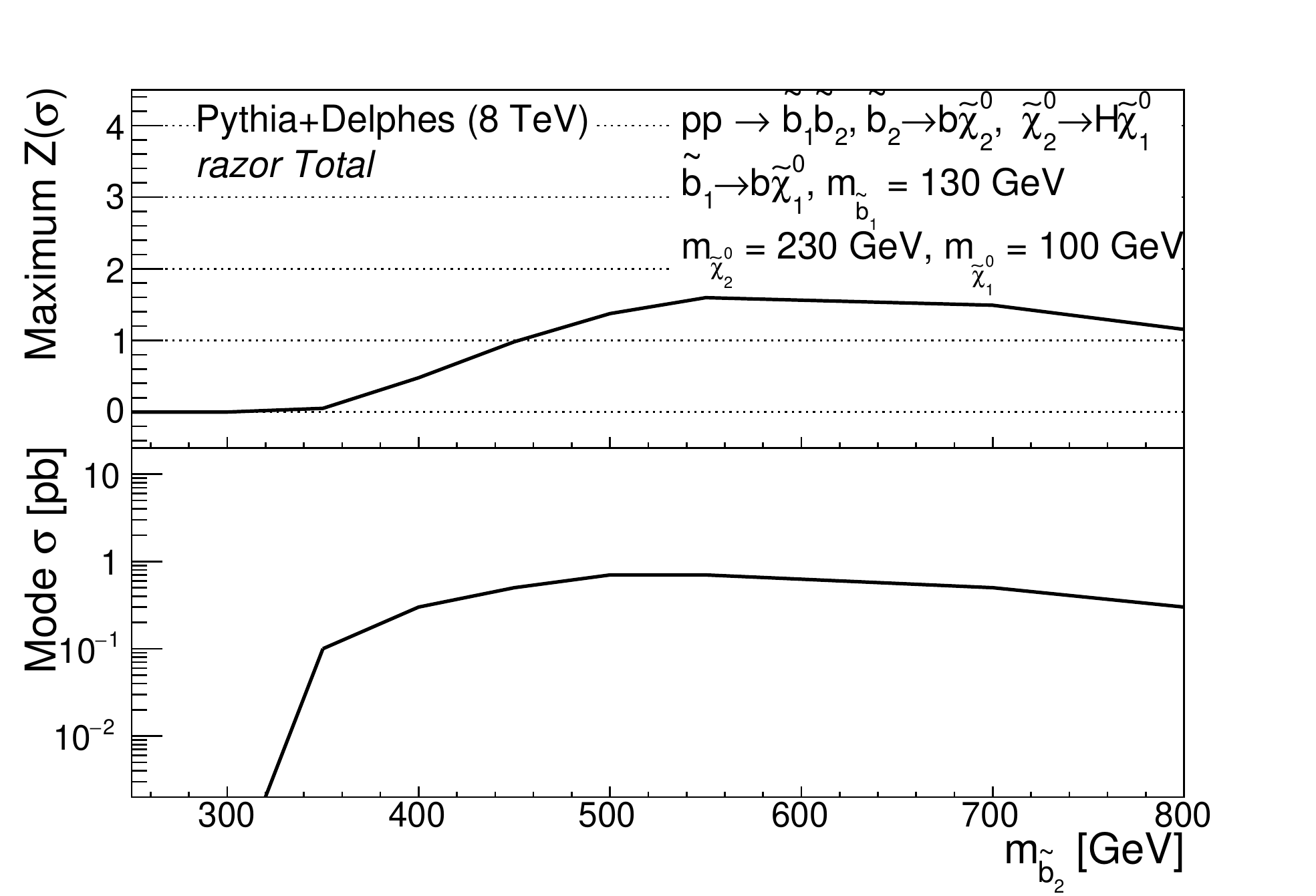}}\\
\subfigure{\includegraphics[width=0.45\textwidth]{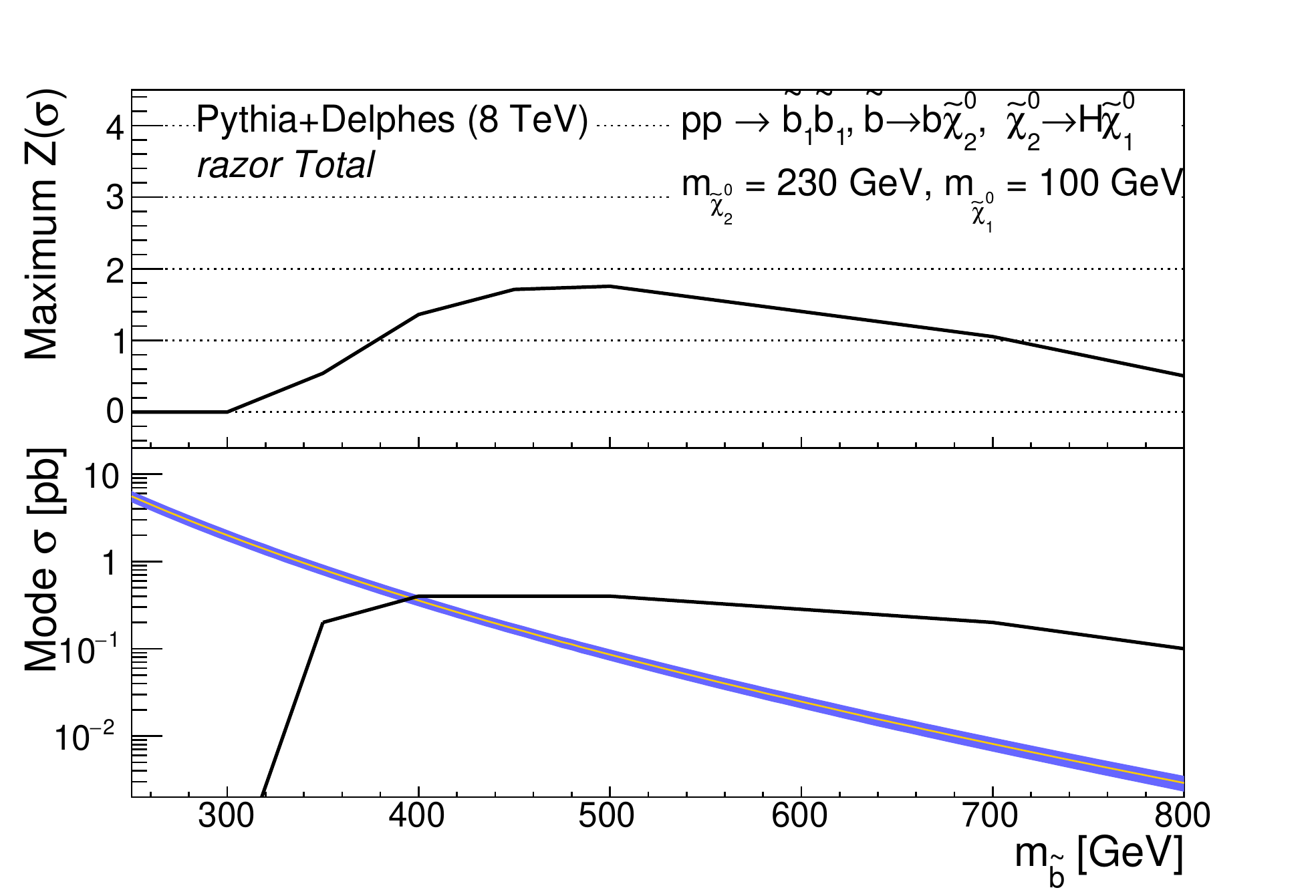}}
\caption{\label{fig:T21bHT2bH1dSignif} (Top) The maximum significance $Z(\sigma)$ for a
  given $m_{\sbottom_2}$ in the top panel and the ``best fit'' signal cross
  section $\sigma$ in the bottom panel for model A.  (Bottom) The maximum significance $Z(\sigma)$ for a
  given $m_{\sbottom_1}$ in the top panel and the ``best fit'' signal cross
  section $\sigma$ in the bottom panel for model B.  Note, these scans assume
  $m_{\chiz_1}=100\GeV$, $m_{\chiz_2}=230\GeV$, and for model A $m_{\sbottom_1}=130$\GeV.}
\end{figure}

Figures~\ref{fig:T21bHT2bHExpObs500800}-\ref{fig:T21bHT2bH1dSignif}
contain the results of the reinterpretation of the CMS data for both models.
To show how well signal model A agrees with the excess observed by CMS,
Fig.~\ref{fig:T21bHT2bHExpObs500800} (top) displays the expected SM background
distribution and uncertainty taken from the CMS result compared to the distribution of the
signal events for $m_{\sbottom_2}=500\GeV$ and
$m_{\sbottom_2}=800\GeV$, with other mass parameters set as
$m_{\sbottom_2}=130\GeV$, $m_{\chiztwo}=230\GeV$, and
$m_{\chizone}=100\GeV$. The bin numbers correspond to the order of the
signal regions in the yield tables in Ref.~\cite{RazorHgaga} and are
reproduced in Tab.~\ref{tab:bins}.
\begin{table}[htb] \centering
\caption{\label{tab:bins}\texttt{HighRes} bin numbering scheme as in Ref.~\cite{RazorHgaga}.}
\begin{tabular}{c|cc}
\hline\hline
Bin &$\MR$ range & $\Rtwo$ range \\
\hline
 0 & $ [150, 250]$ & $[0.00, 0.05]$\\
 1 & $[150, 250]$ & $[0.05, 0.10]$\\
 2 & $[150, 250]$ & $[0.10, 0.15]$\\
 3 &  $[150, 250]$ & $[0.15, 1.00]$\\
 4 &  $[250, 400$ & $[0.00, 0.05]$\\
 5 &  $[250, 400]$ & $[0.05, 0.10]$\\
 6 &  $[250, 400]$ & $[0.10, 1.00]$\\
 7 & $[400, 1400]$ & $[0.00, 0.05]$\\
8 &  $[400, 1400]$ & $[0.05, 1.00]$\\
 9 & $[1400, 3000]$ & $[0.00, 1.00]$\\
\hline\hline
\end{tabular}
\end{table}
The normalization for each signal model is taken from the mode (i.e. ``best-fit'') signal cross section of the posterior density in the
\texttt{HighRes} box. Fig.~\ref{fig:T21bHT2bH1dLimit} (top), shows the
95\% CL combined upper limit on the cross section for model A. Finally,
Fig.~\ref{fig:T21bHT2bH1dSignif} (top) shows the maximum significance
$Z$ as well as the best fit signal cross section for model A as a function of $m_{\sbottom_2}$.

The bottom of Fig.~\ref{fig:T21bHT2bHExpObs500800}-\ref{fig:T21bHT2bH1dSignif} are
the analogous results for model B. The chosen model B mass points in Fig.~\ref{fig:T21bHT2bHExpObs500800} are
$m_{\sbottom_1}=500\GeV$ or $m_{\sbottom_1}=800\GeV$, $m_{\chiztwo}=230\GeV$, and
$m_{\chizone}=100\GeV$. The limit and significance scans in
Fig.~\ref{fig:T21bHT2bH1dLimit}~and~\ref{fig:T21bHT2bH1dSignif} are
performed as a function of the $\sbottom_1$ mass. For model B, we also
compare both the excluded cross section at 95\% CL and the best-fit cross section
as a function of the $\sbottom_1$ mass to the NLO$+$NLL predicted cross section at
$\sqrt{s}=8\TeV$~\cite{NLONLL1,NLONLL2,NLONLL3,NLONLL4,NLONLL5,Borschensky:2014cia}. We
find the $8\TeV$ data excludes bottom squark pair prodction below $m_{\sbottom_1}=330\GeV$ for
the chosen neutralino masses of $m_{\chiztwo}=230\GeV$ and
$m_{\chizone}=100\GeV$. More interestingly, the largest combined significance
is $1.8\sigma$ for $m_{\sbottom_1}=500\GeV$ and the best-fit cross section
is $0.4\unit{pb}$, which is of the same order of magnitude as the
predicted cross section.

\section{Discussion and summary}
\label{sec:conclusionspheno}

In this paper, we proposed two simplified models of bottom squark
pair production for use in the interpretation of an excess observed
by CMS in a search for SUSY in $\PH+$jets events using razor variables at $\sqrt{s}=8\TeV$~\cite{RazorHgaga}. In model A, we considered the
asymmetric production of a $\sbottom_2
\sbottom_1$ pair, with the $\sbottom_1\to\chizone$, $\sbottom_2\to\cPqb
\chiztwo$, and $\chiztwo \to H \chizone$, where $\chizone$ is a
neutralino LSP and we fix the mass splitting $m_{\chiztwo}-m_{\chizone}=130\GeV$. In model B, we considered the symmetric production of a
$\sbottom_1\sbottom_1$ pair, with $\sbottom_1 \to \cPqb \chiztwo$,
$\chiztwo \to \PH \chizone$, and
$m_{\chiztwo}-m_{\chizone}=130\GeV$. 

We scanned the bottom squark masses
for a fixed LSP mass of $m_{\chizone}=100\GeV$ for both models and
quantified the agreement with the data. We found
the excess observed in data is broadly consistent with both models,
with the largest signal significance being $1.8\sigma$
corresponding to model B with $m_{\sbottom_1}=500\GeV$,
$m_{\chiztwo}=230\GeV$, and $m_{\chizone}=100\GeV$. Following this study, model B used by the CMS collaboration to interpret the
results of the updated 13\TeV search for SUSY in the same
channel~\cite{CMS-PAS-SUS-16-012}, which also exhibits an excess
possibly consistent with the model.

\section{Acknowledgements}
This work is funded by the California Institute of Technology High Energy Physics under Contract DE-SC0011925 with the United States Department of Energy.  The authors are grateful to the international  collaborations  of  scientists whose work resulted in the discovery of the Higgs boson at the LHC in 2012 and especially the SUSY searches group of the CMS collaboration.  The work is motivated by results obtained as part of the  Caltech PhD thesis of Alex Mott \cite{alexthesis}. 

\bibliography{RazorGaGaPaper8TeV}{}
\bibliographystyle{unsrt}

\end{document}